\documentclass[12pt]{iopart}

\usepackage{graphicx}
\begin{document}
\title[]{Quantum correlations in Ising-XYZ diamond chain structure under an external magnetic field}
\author{E. Faizi, H. Eftekhari}
\address{Physics Department, Azarbaijan shahid madani university 53714-161, Tabriz, Iran}
\ead{efaizi@azaruniv.edu, h.eftekhari@azaruniv.edu}
\begin{abstract}
In this paper, we consider an entangled Ising-XYZ diamond chain structure. Quantum correlations for this model is investigated by using of quantum discord and trace distance discord. Quantum correlations is obtained for different values of the anisotropy parameter, magnetic field and temperature. By comparison between quantum correlations, we show trace distance discord is always larger than quantum discord. Finally, some novel effects, such as increasing the quantum correlations with temperature and constructive role of anisotropy parameter which may paly to the quantum correlations were observed.
\end{abstract}

\section{Introduction}
Quantum discord (QD) has received much attention, since it was proposed by Ollivier and Zurek \cite{H. Ollivier}. The quantum discord which measures a more generic kind of quantum correlation, is known to have nonzero values even for separable mixed states \cite{O. Harold}. QD is made on the fact that two classical equivalent ways of defining the mutual information are not equivalent in the quantum domain. In general, it is to some extent hard to calculate QD and analytical solutions can hardly be found except for some certain cases, such as the so-called X states \cite{M. Ali}. The difficulty of finding analytical solutions for QD led Dakic et al. to introduce a geometric measure of QD \cite{B. Dakic}, which determines the amount of quantum correlations of a state in terms of its minimal Hilbert-Schmidt distance from the set of classical states. The calculation of this measure needs an easier minimization process, which is obtainable analytically for general two-qubit states \cite{B. Dakic} as well as for arbitrary bipartite states \cite{S. Luo, A. S. M. Hassan}. Such a distance is well-known not to contractive under trace-preserving quantum channels \cite{I. Bengtsson, M. Ozawa}. This naturally leads to a redefinition of the geometric discord in terms of a metric that fulfills the contractivity property. One such metric is the trace distance \cite{M. A. Nielsen, M. B. Ruskai}, which uses the Schatten one-norm (or trace norm). In this paper, we refer to this particular geometric measure as trace distance discord (TDD).\\
In last years, with the development of quantum information and quantum computation, spin systems have an extensive application in the field of quantum information \cite{X. S. Ma, K. T. Guo}. Heisenberg model, as the simplest spin chain,  has been researched in many fields of quantum information and computation. It can be implemented in many physical systems such as quantum dot system \cite{B. E. Kane}, nucleus system \cite{R. Vrijen}, electronic spin system \cite{A. Sorensen} and optical lattices system \cite{D. A. Lidar} and so on. Quantum correlation in the Heisenberg model is like a bridge between quantum information and condensed matter physics. \\
Motivated by real materials like $Cu_3(CO_3)_2(OH)_2$ known as azurite, which is an exciting quantum anti-ferromagnetic model characterized by Heisenberg model on generalized diamond chain. Honecker et al. investigated the dynamic and thermodynamic properties for this model \cite{A. Honecker}. Furthermore, the thermodynamics of the Ising-Heisenberg model on diamond-like chain was also vastly discussed (see \cite{L. Canova, J. S. Valverde}). The motivation to research the Ising-XYZ diamond chain model is based in some recent works. Lately, the properties of thermal entanglement have been studied in the Ising-XXZ model on diamond chain \cite{O. Rojas}, and in the Ising-XYZ model on diamond chain \cite{J. Torrico}. Therefore, it is quite necessary to study the QD of the Ising-XYZ on diamond chain. Also, only a few studies are directed to the relation between QD and TDD in the quantum channels \cite{E. Faizi}. Therefore, here we also compare the QD and TDD and illustrate their different characteristics. This paper is organized as follows: in Sec. 2 we present the Ising-XYZ model on diamond chain. Further, in Sec. 3 we review the definition of QD and TDD. In Sec. 4 we have discussed QD and TDD of Heisenberg reduced density operator of the model. Finally, concluding remarks are given in Sec.5.

\section{Quantum discord and Trace distance discord}
\subsection{QD}
In classical information theory, the total correlation between
two arbitrary parts can be expressed as two kinds of equivalent expressions of mutual  information. In the quantum domain, one of quantum extension of mutual information is equal to total correlation. It can be expressed as:
\begin{eqnarray}I(\rho_{AB})=S(\rho_A)+S(\rho_B)-S(\rho_{AB}),
\end{eqnarray}
where $S(\rho)=-\Tr(\rho\log_2\rho)$  is Von Neumann entropy and $\rho_A(\rho_B)$ is the reduced density operator of the part $A(B)$. The other quantum version of mutual information can be written after a complete set of projection measurements $\{B_k\}$. Since the systems have quantum correlation, the
quantum correlation will unavoidably cause to another system disturbed when we measure one quantum system. Therefore, the two quantum extensions of mutual information are not equal to each other. However the maximum of the second extension can be interpreted as a measure of classical correlations $C(\rho_{AB})$. It can be written as \cite{L. Henderson, N. Li}:
\begin{eqnarray}C(\rho_{AB})\equiv{S(\rho_A)-\min_{B_k}\tilde{S}(\rho_{AB}|\{B_k\})},
\end{eqnarray}
where $\tilde{S}(\rho_{AB}|\{B_k\})=\sum_k{p_k}S(\rho_{AB}^k)$ is the conditional entropy of partition $A$, $\rho_{AB}^k=(I\otimes{B_k})\rho_{AB}(I\otimes{B_k})/p_k$ and $p_k=\Tr[(I\otimes{B_k})\rho_{AB}(I\otimes{B_k})]$. The minimum value in Eq. (2) is due to the complete set of projection measurements $\{B_k\}$. The minimum difference between the two quantum versions of mutual information is equal to QD. It can be expressed as \cite{H. Ollivier}:
\begin{eqnarray}D(\rho_{AB})=I(\rho_{AB})-C(\rho_{AB}).
\end{eqnarray}

\subsection{TDD}
 At first, we briefly review the definition and the general formalism for the trace distance discord. For any bipartite system $AB$ which described by the density operator $\rho$, the
trace distance is defined as the minimal trace distance between $\rho$ and all of the classical-quantum states $\rho_{CQ}$ \cite{F. M. Paula}, namely,
\begin{eqnarray}D_T(\rho)=\min_{\chi\in{\rho_{CQ}}}\|\rho-\chi\|_1,
\end{eqnarray}
where $\|X\|_1=\Tr\sqrt{X^\dagger{X}}$ shows the trace norm (Schatten 1-norm), and $\rho_{CQ}$ is in the following form
\begin{eqnarray}\rho_{CQ}=\sum_ip_i\Pi_i^A\otimes{\rho_i^B}
\end{eqnarray}
which is a linear combination of the tensor products of $\Pi_i^A$ (the orthogonal projection in the Hilbert space $H_A$) and $\rho_i^B$ (an arbitrary density operator in $H_B$, with $\{p_i\}$ being a probability distribution. For the certain case of the two-qubit X states $\rho^X$ whose nonzero element are along only the main diagonal and anti-diagonal \cite{T. Yu}, the
trace distance discord can be expressed analytically \cite{F. Ciccarello}, which has the following compact form
\begin{eqnarray}D_T(\rho^X)=\sqrt{\frac{\gamma_1^2\gamma_{\max}^2-\gamma_2^2\gamma_{\min}^2}{\gamma_{\max}^2-\gamma_{\min}^2+\gamma_1^2-\gamma_2^2}}
\end{eqnarray}
where $\gamma_{1,2}=2(|\rho_{23}|\pm{|\rho_{14}|})$, $\gamma_3=1-2(\rho_{22}+\rho_{33})$, $\gamma_{\max}^2=\max\{\gamma_3^2,\gamma_2^2+x_{A3}^2\}$ and $\gamma_{\min}^2=\min\{\gamma_1^2,\gamma_3^2\}$, with $x_{A3}=2(\rho_{11}+\rho_{22})-1$.

\section{THE CORRELATIONS IN THE ISING-XYZ CHAIN ON DIAMOND CHAIN STRUCTURE}
 Pairwise thermal entanglement in Ising-XYZ diamond chain structure already was discussed in Ref. \cite{J. Torrico}. In this work, we will discuss the quantum discord and trace distance discord for this model. An Ising-XYZ diamond chain structure is schematically illustrated in figure 1. The Ising-XYZ Hamiltonian can be written in following form \cite{J. Torrico}:
\begin{eqnarray}H=-\sum_{i=1}^N[J(1+\gamma)\sigma_{a,i}^x\sigma_{b,i}^x+J(1-\gamma)\sigma_{a,i}^y\sigma_{b,i}^y+J_z\sigma_{a,i}^z\sigma_{b,i}^z\\\nonumber
+J_0(\sigma_{a,i}^z+\sigma_{b,i}^z)(S_i+S_{i+1})+h(\sigma_{a,i}^z+\sigma_{b,i}^z)+\frac{h}{2}(S_i+S_{i+1})],
\end{eqnarray}
where $\sigma_{a(b)}^\alpha$ are the Pauli matrix with $\alpha=\{x,y,z\}$, and $S$ corresponds to the Ising spins, whereas $\gamma$ is the XY-anisotropy parameter, and $h$ indicates magnetic field.

\begin{figure}
\includegraphics[width=4.6in]{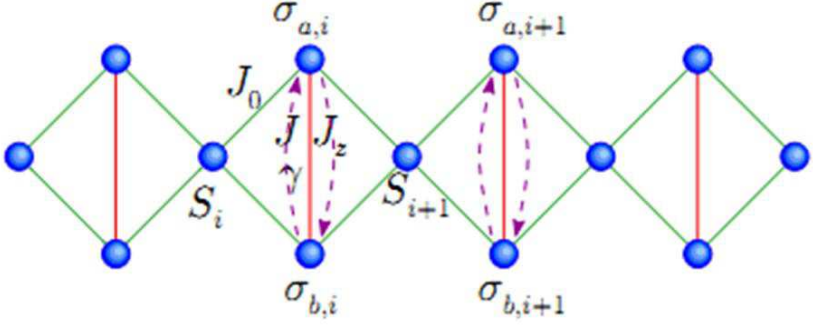}
\caption{Schematic representation of Ising-XYZ chain on diamond structure, $\sigma_{a,i}$ and $\sigma_{b,i}$ are Heisenberg spins, while $S_i$ corresponds to Ising spins.}
\label{fig1}
\end{figure}

For the case of an infinite chain, the reduced density matrix can be expressed as:
\begin{eqnarray}
\rho_X=\left(
\begin{array}{cccccccc}
\rho_{11}&&0&&0&&\rho_{14}\\
0&&\rho_{22}&&\rho_{23}&&0\\
0&&\rho_{23}&&\rho_{33}&&0\\
\rho_{14}&&0&&0&&\rho_{44}\\
\end{array}
\right),
\end{eqnarray}

In which the elements of density matrix could be expressed in terms of the correlation function between two entangled particles \cite{L. Amico},
\begin{eqnarray}\rho_{11}=\frac{1}{4}+\langle{\sigma_a^z\sigma_b^z}\rangle+\langle{\sigma_a^z}\rangle,\\\nonumber
\rho_{22}=\rho_{33}=\frac{1}{4}-\langle{\sigma_a^z\sigma_b^z}\rangle,\\\nonumber
\rho_{44}=\frac{1}{4}+\langle{\sigma_a^z\sigma_b^z}\rangle-\langle{\sigma_a^z}\rangle,\\\nonumber
\rho_{14}=\langle{\sigma_a^x\sigma_b^x}\rangle-\langle{\sigma_a^y\sigma_b^y}\rangle,\\\nonumber
\rho_{23}=\langle{\sigma_a^x\sigma_b^x}\rangle+\langle{\sigma_a^y\sigma_b^y}\rangle,
\end{eqnarray}
At finite temperature, each expected values become temperature dependent quantities which are in the following form: \cite{J. Torrico}:
\begin{eqnarray}\langle{\sigma_a^x\sigma_b^x}\rangle=\e^{\beta\frac{h}{2}}\frac{\frac{\Delta(1)}{2}\e^{-\beta\frac{2h+J_z}{4}}\sinh(\beta\frac{J}{2})+\frac{J\gamma}{4}\e^{\beta\frac{J_z}{4}}\sinh(\beta\Delta(1))}{\Delta(1)\lambda_+}\\\nonumber
\langle{\sigma_a^y\sigma_b^y}\rangle=\e^{\beta\frac{h}{2}}\frac{\frac{\Delta(1)}{2}\e^{-\beta\frac{J_z}{4}}\sinh(\beta\frac{J}{2})-\frac{J\gamma}{4}\e^{\beta\frac{J_z}{4}}\sinh(\beta\Delta(1))}{\Delta(1)\lambda_+}\\\nonumber
\langle{\sigma_a^z\sigma_b^z}\rangle=\e^{\beta\frac{h}{2}}\frac{\e^{\beta\frac{J_z}{4}}\cosh(\beta\frac{J}{2})-\e^{-\beta\frac{J_z}{4}}\cosh(\beta\Delta(1))}{2\lambda_+}\\\nonumber
\langle{\sigma_a^z}\rangle=\e^{\beta\frac{2h+J_z}{4}}\sinh(\beta\Delta(1))\frac{J_0+h}{\Delta(1)\lambda_+}
\end{eqnarray}
wherein
\begin{eqnarray}\Delta(\mu)=\sqrt{(h+{\mu}J_0)^2+\frac{1}{4}J^2\gamma^2}
\end{eqnarray}
\begin{eqnarray}\lambda_+=\omega(2)+\omega(-2)+\sqrt{(\omega(2)-\omega(-2))^2+4\omega(0)^2}
\end{eqnarray}
in which
\begin{eqnarray}\omega(\mu)=2\e^{\frac{\beta\mu{h}}{2}}[\e^{-\frac{{\beta}J_z}{4}}\cosh(\frac{{\beta}J}{2})+\e^{\frac{{\beta}J_z}{4}}\cosh(\beta\Delta(\mu))]
\end{eqnarray}
where $\beta=1/k_BT$, with with $k_B$ being the Boltzmann constant and $T$ is the absolute temperature. For simplicity, we write $k=1$.\\
 Finding analytical solution for quantum discord is difficult to some extent. Nevertheless, for the X states described by the density matrix equation (8), one can find the explicit expression of quantum discord as \cite{F. F. Fanchini}:
\begin{eqnarray}D(\rho)=\min\{D_1,D_2\}
\end{eqnarray}
where
\begin{eqnarray}D_1=S(\rho^A)-S(\rho^{AB})-\rho_{11}\log_2(\frac{\rho_{11}}{\rho_{11}+\rho_{22}})-\rho_{22}\log_2(\frac{\rho_{22}}{\rho_{11}+\rho_{22}})\\\nonumber
-\rho_{44}\log_2(\frac{\rho_{44}}{\rho_{22}+\rho_{44}})-\rho_{22}\log_2(\frac{\rho_{22}}{\rho_{22}+\rho_{44}})
\end{eqnarray}
and
\begin{eqnarray}D_2=S(\rho^A)-S(\rho^{AB})-\Delta_+\log_2\Delta_+-\Delta_-\log_2\Delta_-
\end{eqnarray}
with $\Delta_{\pm}=\frac{1}{2}(1\pm{\Gamma})$ and $\Gamma^2=(\rho_{11}-\rho_{44})^2+4(|\rho_{14}|+|\rho_{23}|)^2$.
Finally, according to Eq. (6), TDD can be obtained easily.

\begin{figure}
\includegraphics[width=2.8in]{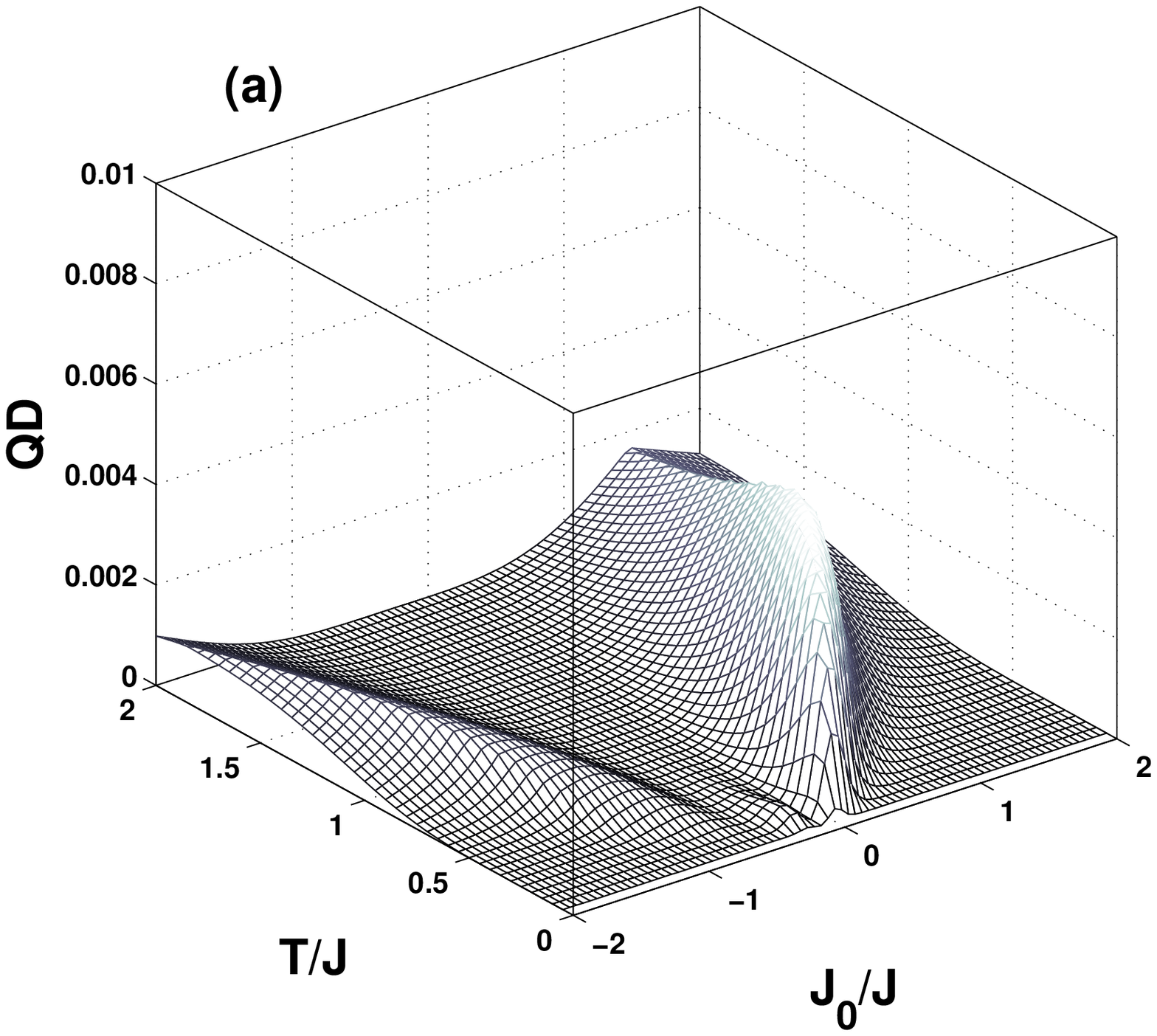}
\includegraphics[width=2.8in]{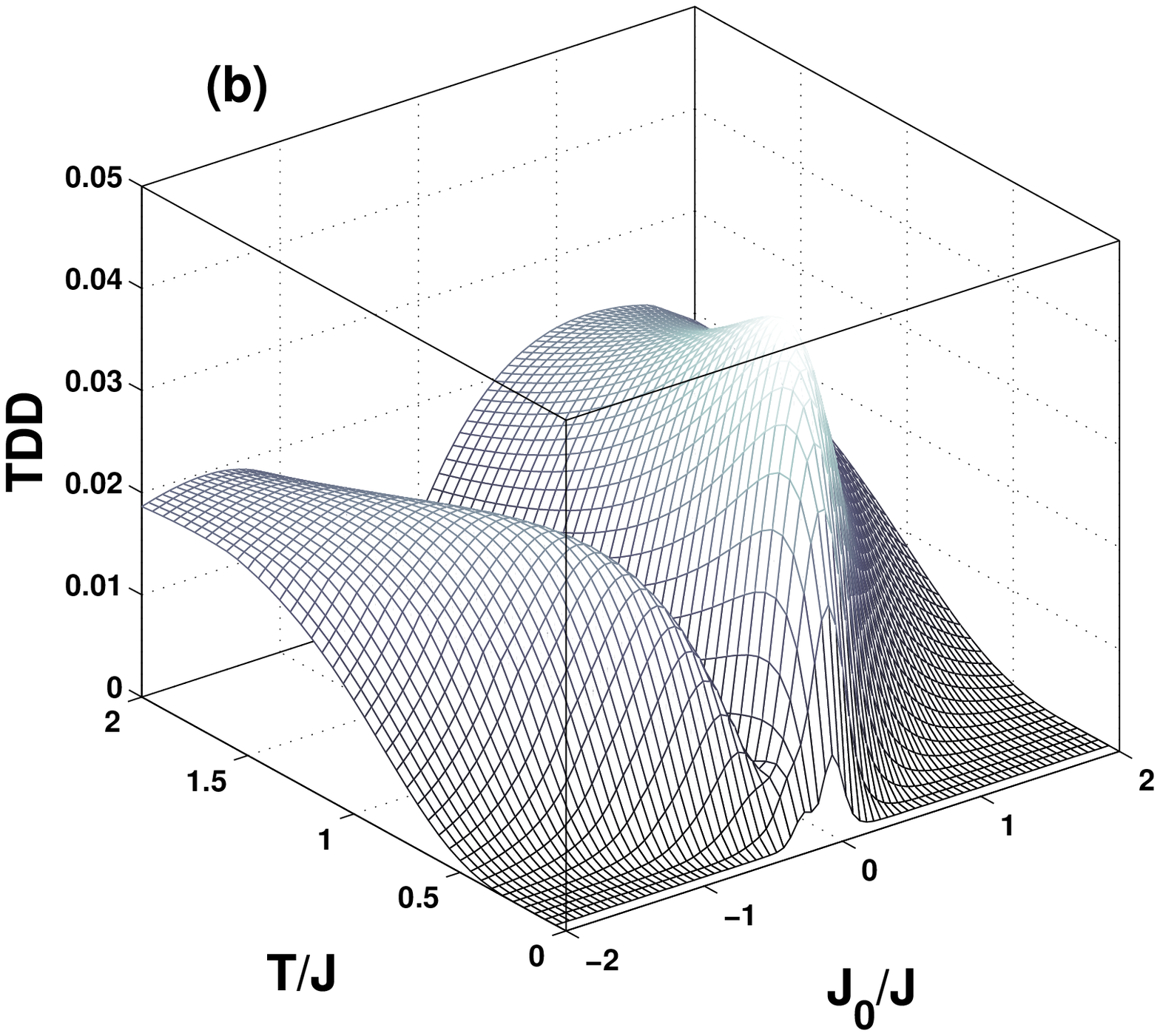}
\includegraphics[width=2.8in]{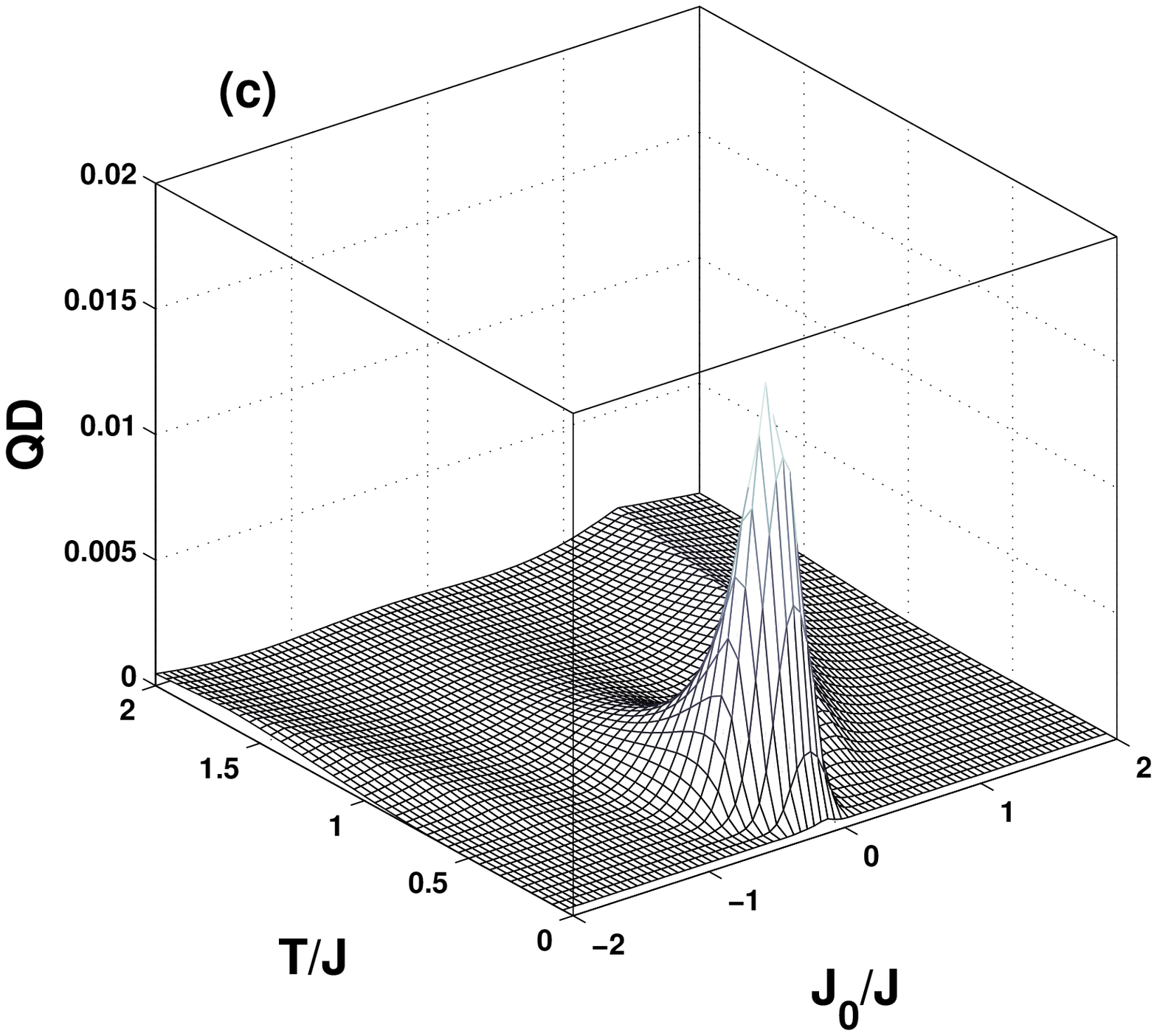}
\includegraphics[width=2.8in]{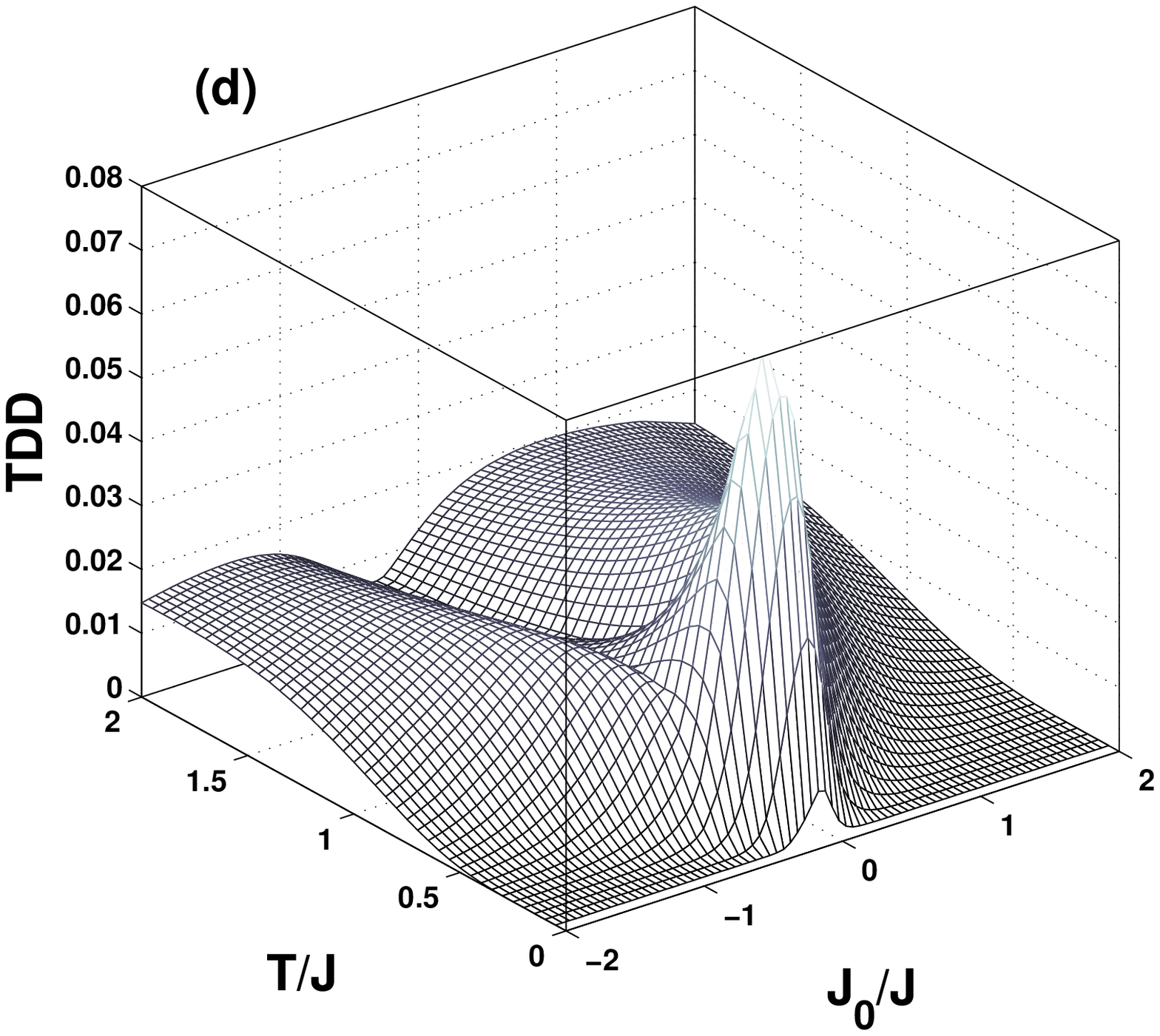}
\caption{Quantum correlations as a function of $J_0/J$ and $T/J$, with ${J_z}/{J}=0$, $\gamma=0.95$, ${h}/{J}=0.27$ (a) and (b); ${J_z}/{J}=0.3$, $\gamma=0.6$, ${h}/{J}=0.35$ (c) and (d).}
\label{fig2}
\end{figure}

\begin{figure}
\includegraphics[width=2.8in]{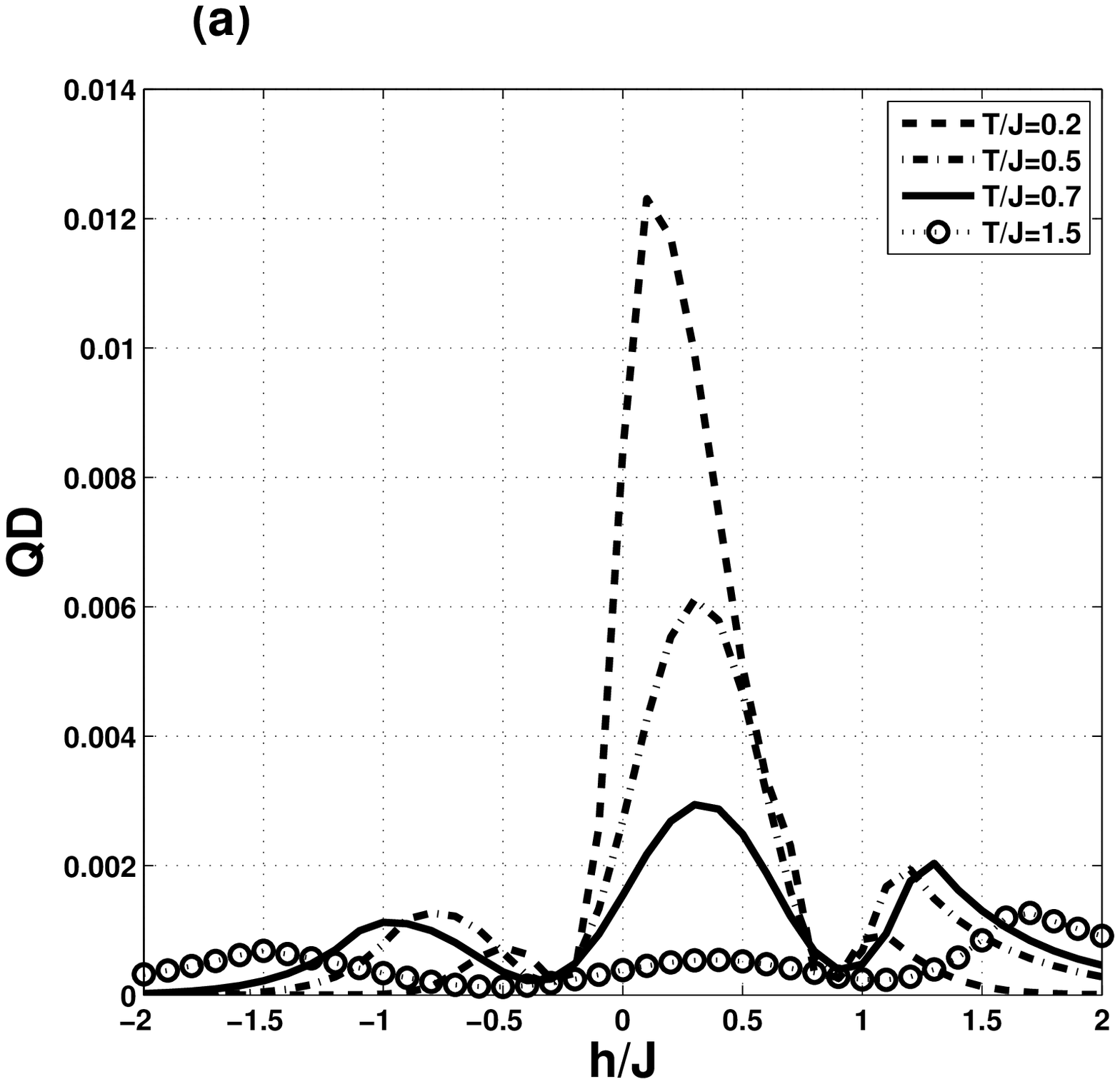}
\includegraphics[width=2.8in]{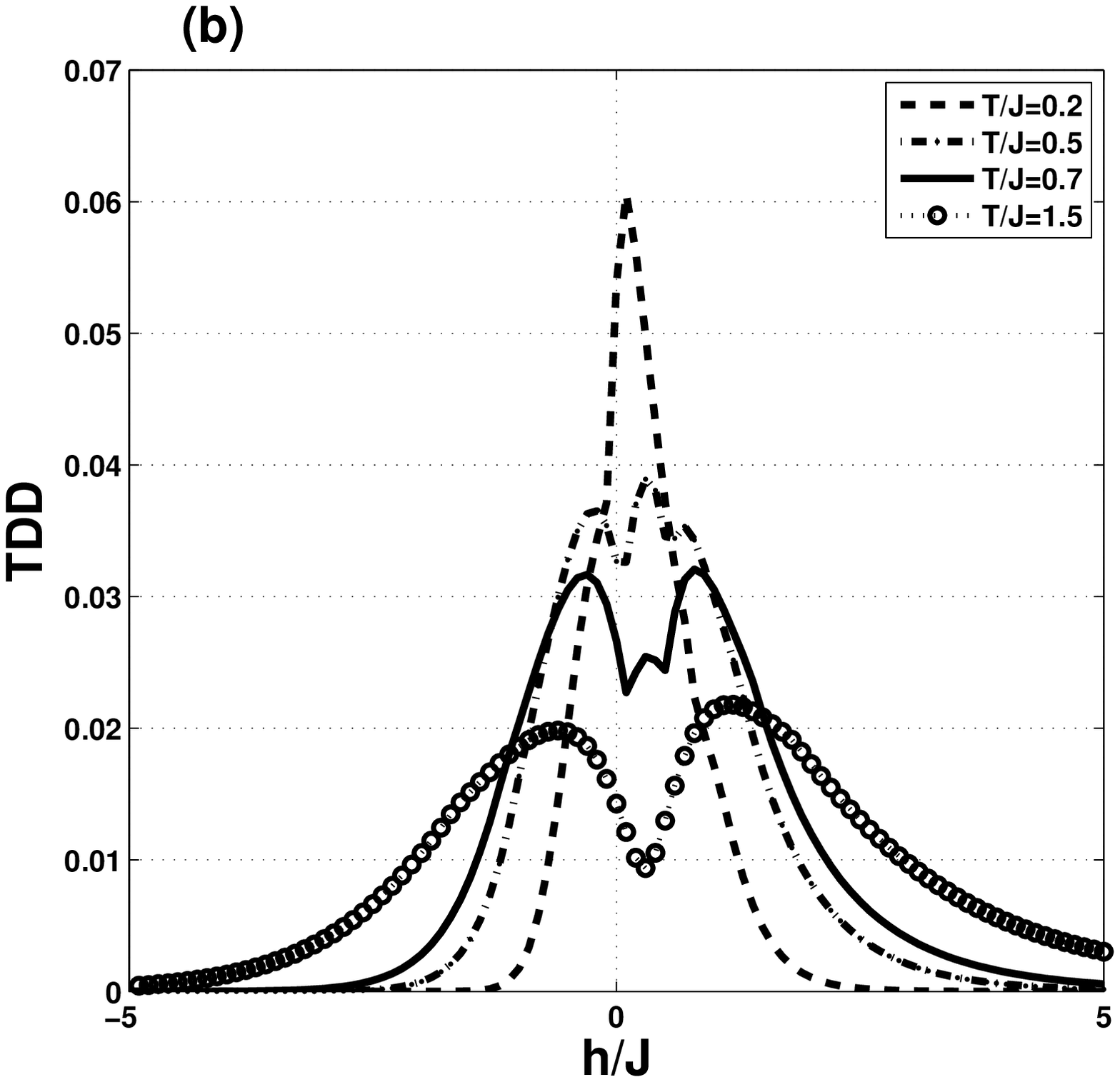}
\caption{QD (a), TDD (b) as a function of $h/J$ with $\gamma=0.5$, ${J_0}/{J}=-0.3$, ${J_z}/{J}=0.3$ and different values of temperature $T/J$: $T/J=0.2$ (dash line), $T/J=0.5$ (dash-dot line), $T/J=0.7$ (solid line), $T/J=1.5$ (circle line).}
\label{fig3}
\end{figure}

\begin{figure}
\includegraphics[width=2.8in]{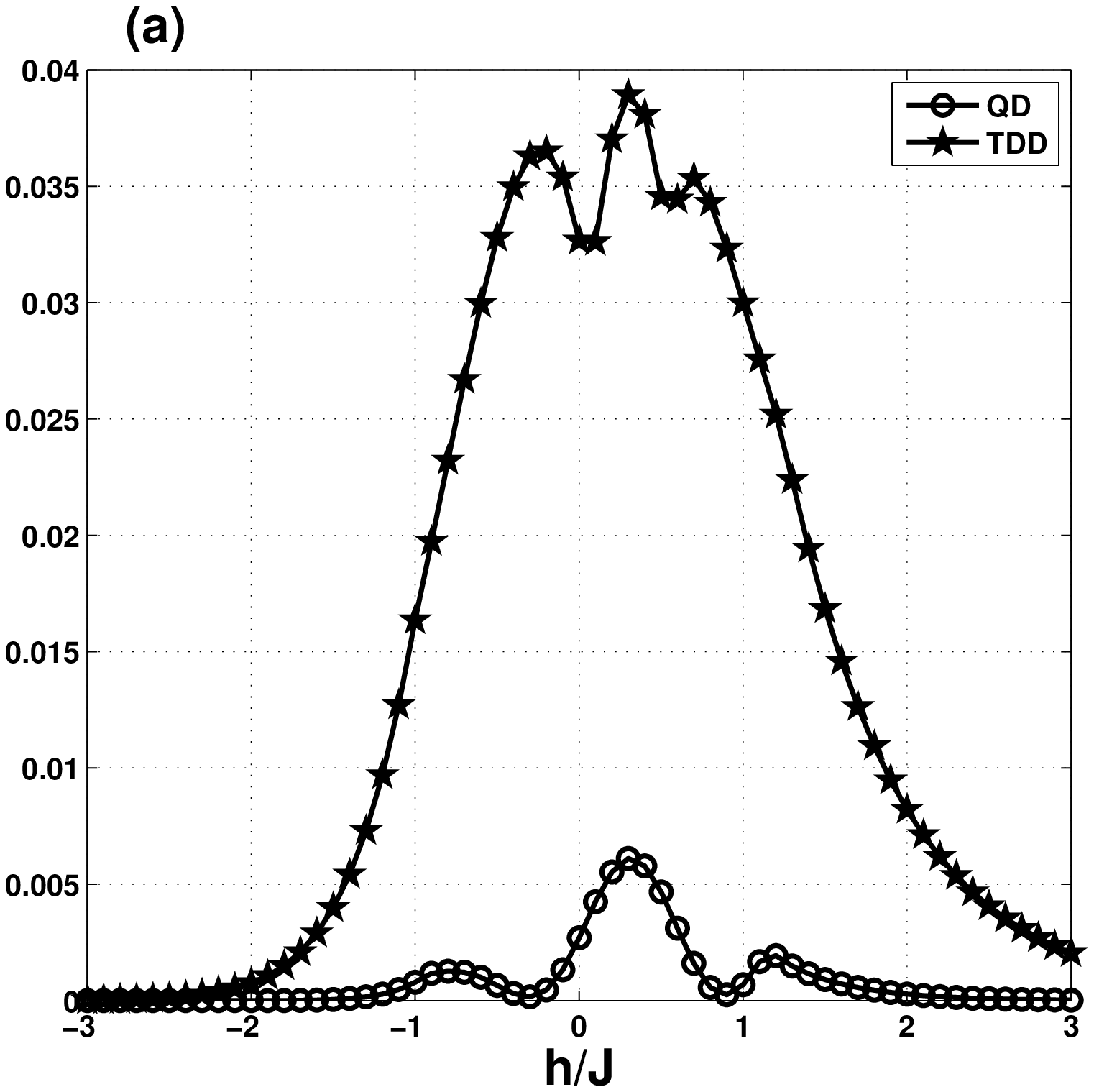}
\includegraphics[width=2.8in]{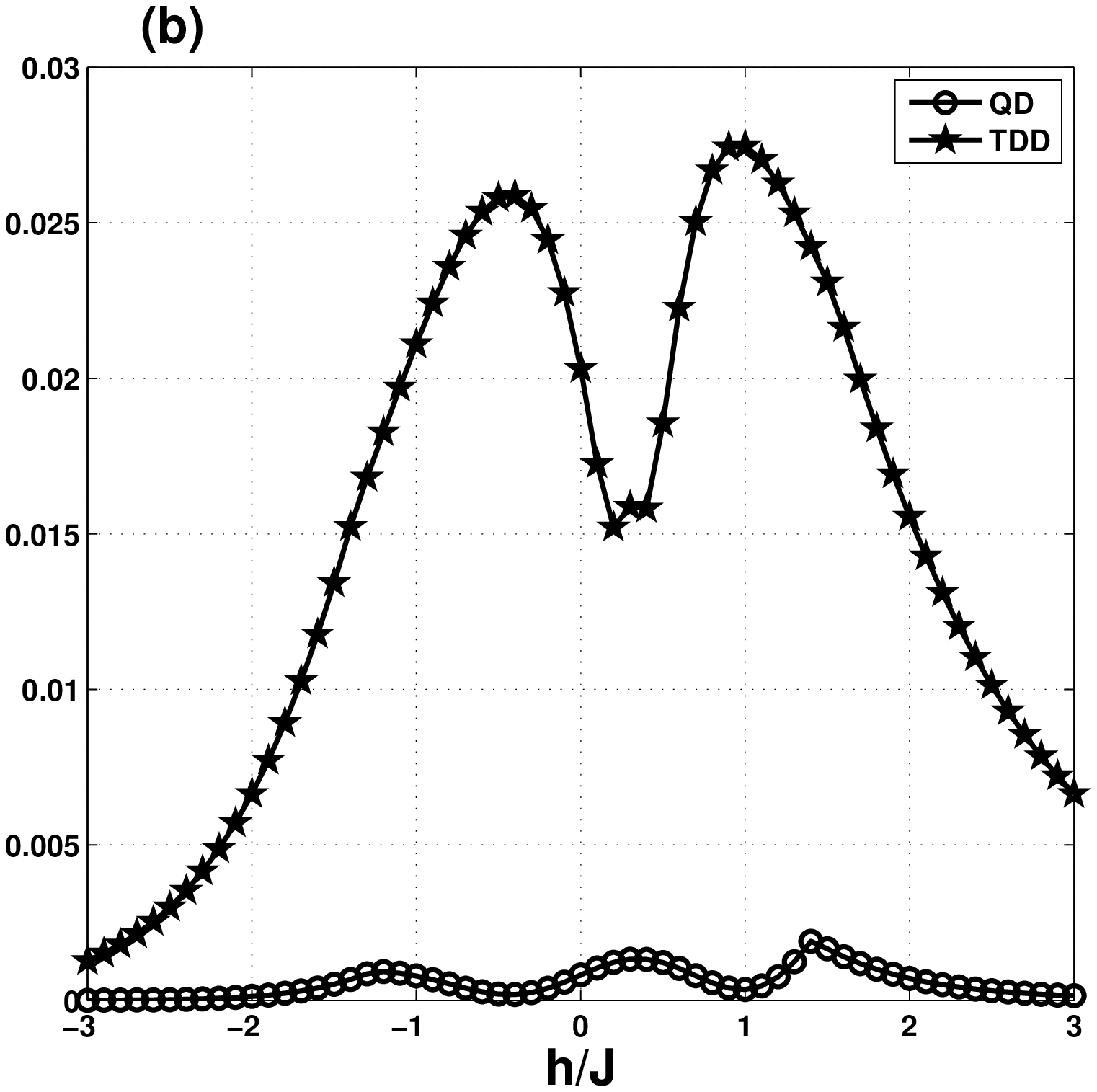}
\caption{Comparison of QD and TDD versus $h/J$, with $\gamma=0.5$, ${J_0}/{J}=-0.3$, ${J_z}/{J}=0.3$; $T/J=0.5$(a), $T/J=1$ (b).}
\label{fig4}
\end{figure}

\begin{figure}
\includegraphics[width=2.8in]{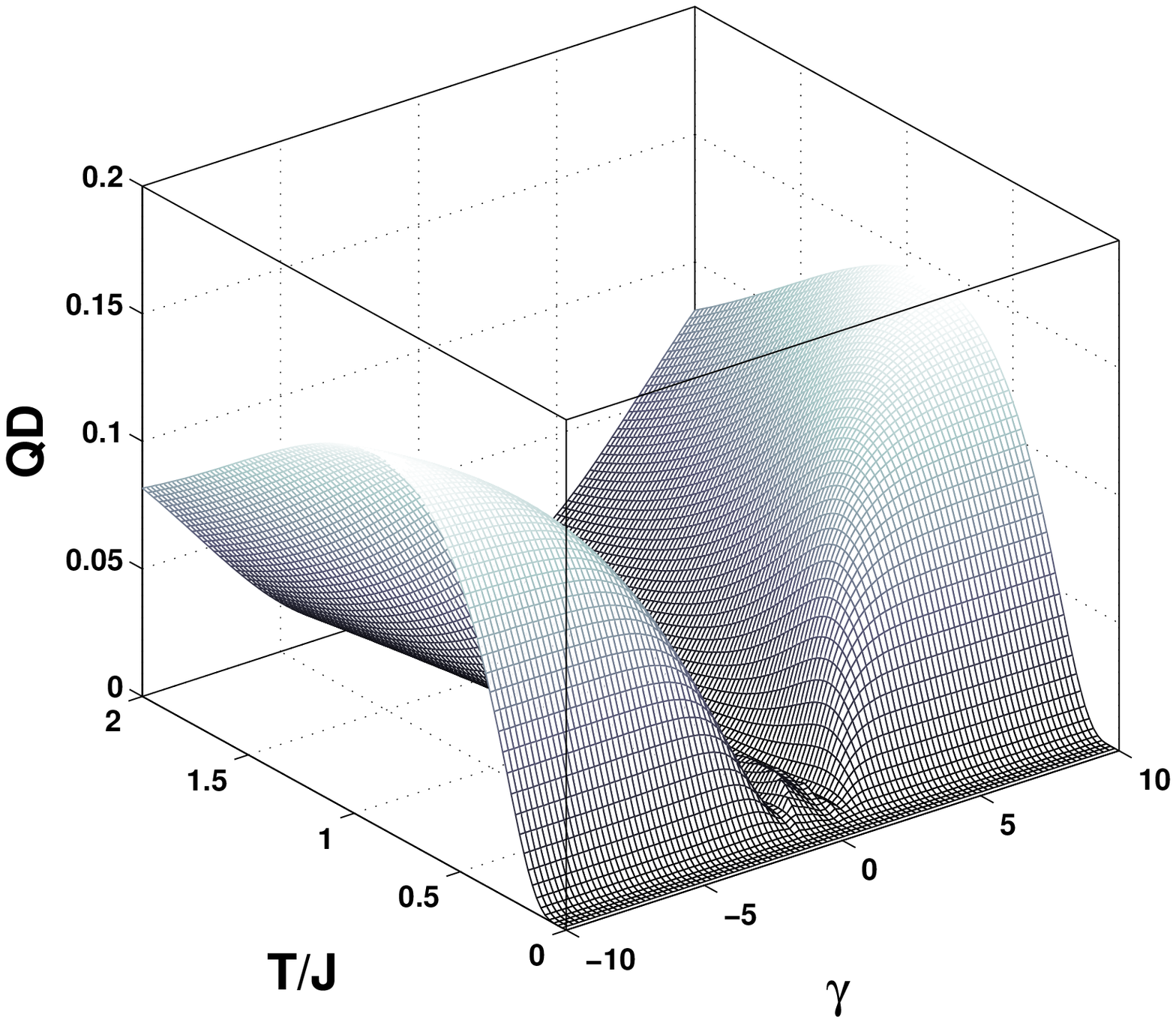}
\includegraphics[width=2.8in]{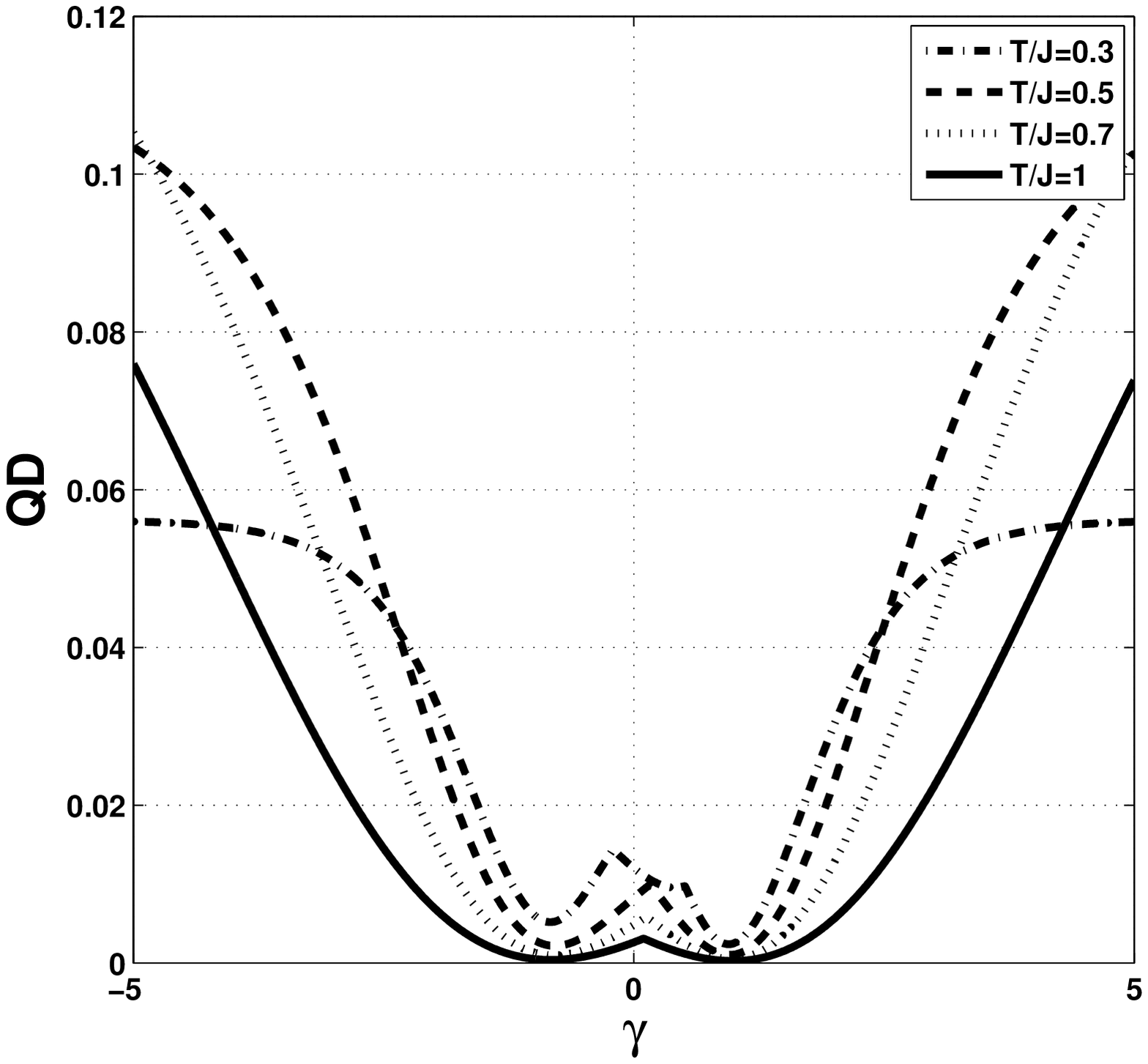}
\includegraphics[width=2.8in]{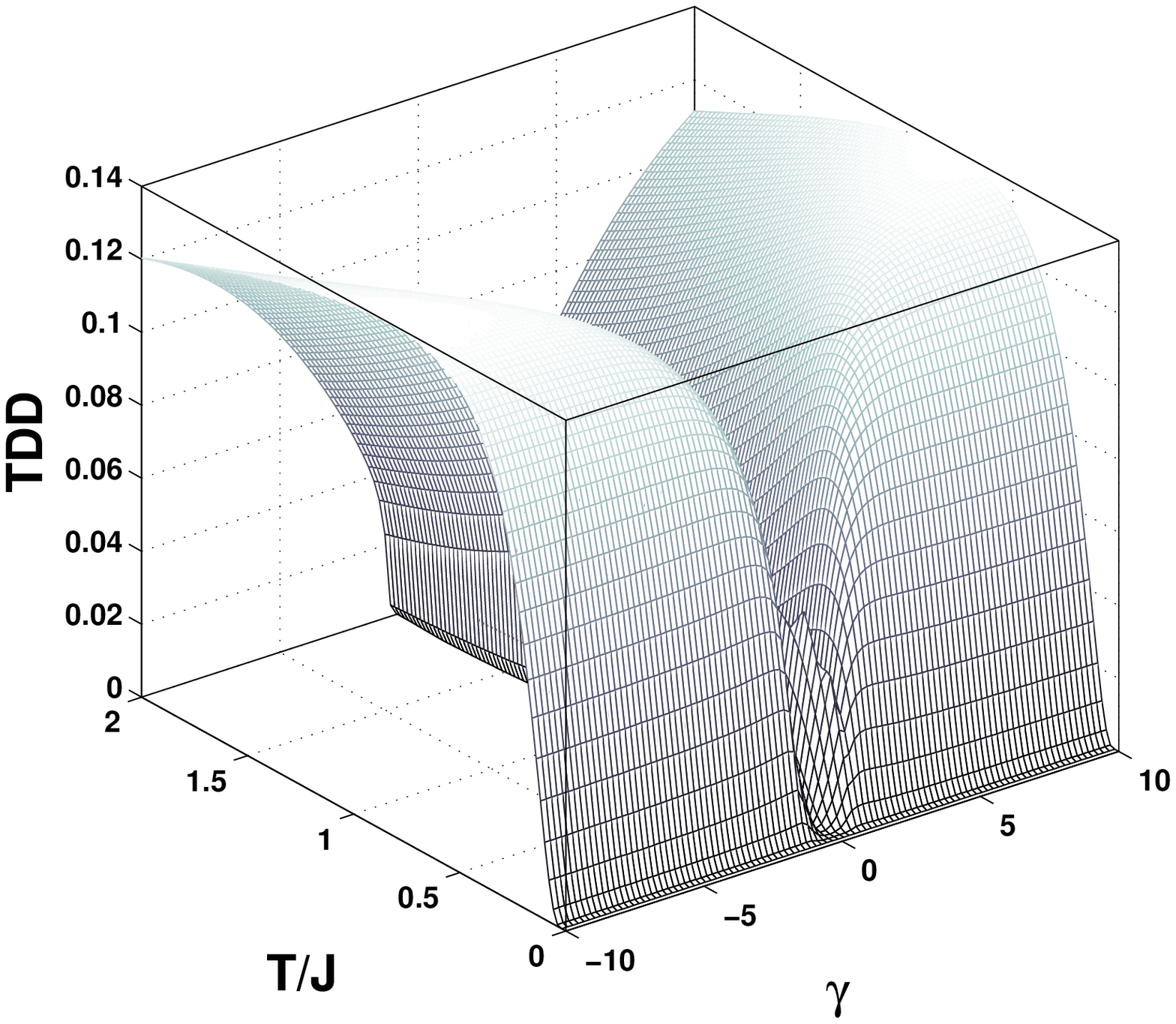}
\includegraphics[width=2.8in]{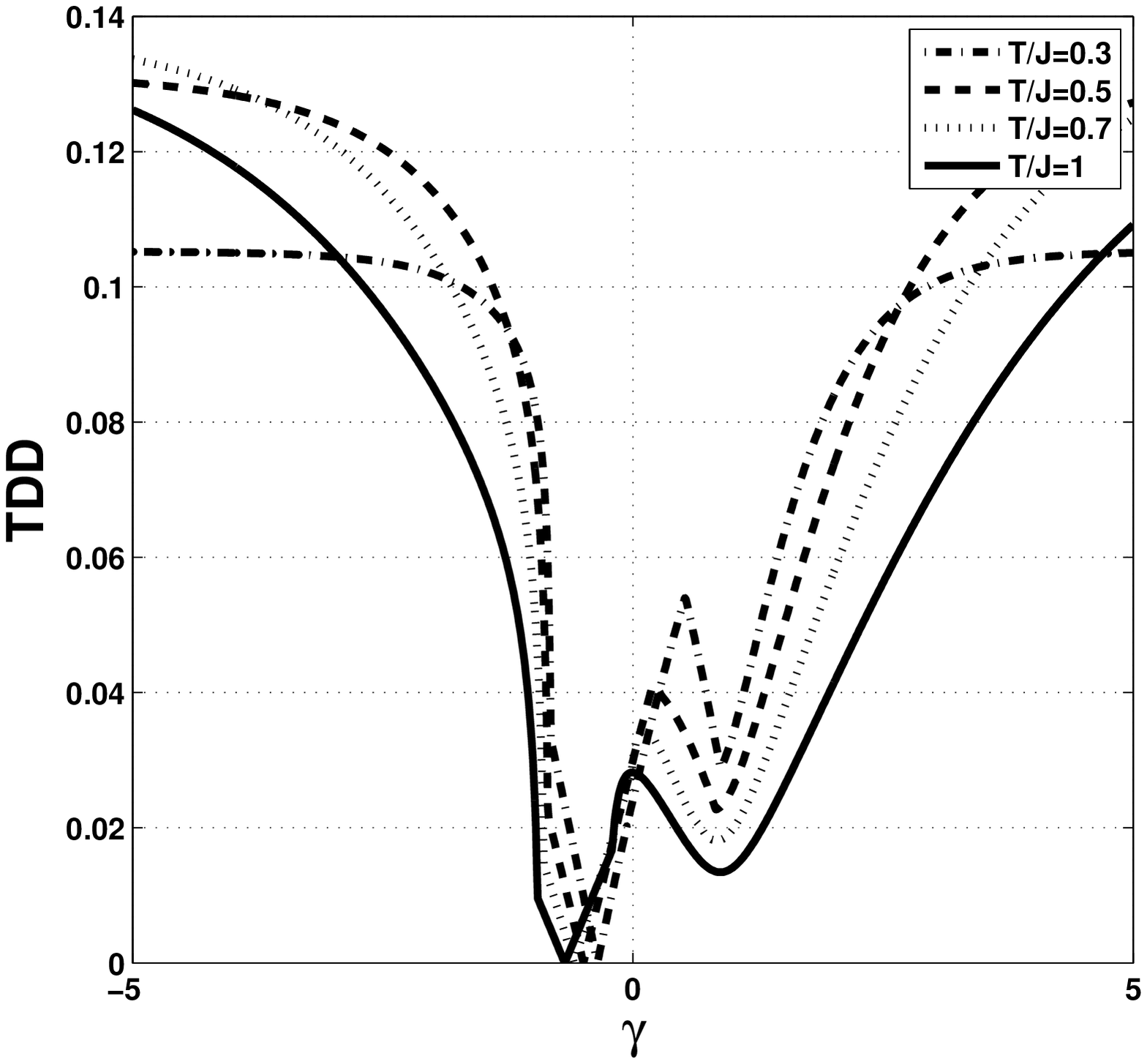}
\caption{Quantum correlations as a function of $\gamma$ and $T/J$ with ${J_0}/{J}=-0.3$, ${J_z}/{J}=0.3$ ${h}/{J}=0.5$.}
\label{fig5}
\end{figure}

According to the Eqs. (6) and (14) QD and TDD can be worked out by numerical calculation. We will now discuss them with the corresponding plots. The QD and TDD both are known to vary with the parameters $J_0$, $h$, $\gamma$, $J_z$ and T. By fixing some of the parameters we can analyze the roles that the others play in QD and TDD.\\
 In Fig. 2 quantum correlations are plotted versus $J_0/J$ and $T/J$, for fixed parameters $J_0/J=0$, $\gamma=0.95$, $h/J=0.27$ in panels (a) and (b) and for fixed parameters $J_0/J=0.3$, $\gamma=0.6$, $h/J=0.35$ in panels (c) and (d). We can see from Fig. 2, both QD and TDD increase from zero to some peak values in the beginning and then decrease with temperature increasing. Note that the peak value depend on $J_0/J$, so that smaller $J_0/J$ have higher peak values. Moreover, after reaching to peak values QD and TDD decrease more slowly for larger value of $J_0/J$, as temperature goes up, than smaller value of $J_0/J$. The distinct difference between QD and TDD is: TDD has larger value than TDD. That the QD and TDD do not monotonically decline with increasing temperature shows the more correlated low-lying excited states in some regions \cite{J. Maziero}. In addition, pairwise thermal entanglement of Ising-XYZ diamond chain model has been considered in Ref. \cite{J. Torrico}. The authors found that thermal entanglement vanishes as for as the temperature increases, while we find from Fig. 2 that, QD and TDD do not vanish up to infinite temperature. It indicates that, both of the QD and TDD are more general quantum correlations than entanglement.\\
 We now turn to characterize the dependence of QD and TDD on magnetic field.
 Fig. 3 shows the behavior of QD and TDD versus magnetic field and different temperature at a fixed value of $\gamma=0.5$ and $J_0/J=-0.3$, $J_z/J=0.3$. For the QD case, as the temperature is low ($T/J=0.2$) there are three sharp peaks (Fig. 3(a)). As we increase temperature, the middle peak gets shorter. Compared with QD, we can see from Fig. 3(b) the behavior of TDD is different from QD to some extent. As the temperature is low ($T/J=0.2$), we can see obviously only one peak. As we increase temperature ($T/J=0.5$), one can see that the one peak evolve into three peaks. That is to say, the left and right peaks appear. When $T/J=0.7$ the middle peak get shorter. As the temperature is further increased, for example $T/J=1.5$, the middle peak disappears and we can see two sharp peaks. Also, QD and TDD finally vanish as magnetic field increases, namely strong magnetic field is harmful to quantum correlation. In Fig. 4 the behaviors of two quantities versus magnetic field at two certain temperature ($T=0.5$ Fig. 4(a) and $T=1$ Fig. 4(b)) are compared. We can known that the two quantities are not symmetric with respect to the zero magnetic field, moreover the QD has always lower value than TDD at the temperature both $T=0.5$ and $T=1$.\\
  We now move to investigate how the quantum correlations behaves as we change the anisotropy parameter at finite temperature. The QD and TDD as a function of $\gamma$ and $T/J$ for fixed values of $h/J=0.5$, $J_0/J=-0.3$ and $J_z/J=0.3$ is shown in Fig. 5. It is seen that QD and TDD are not a symmetric function about zero anisotropy parameter $\gamma$. We can see that QD and TDD evolve similar but not the same which can be seen clearly from Fig. 5. As increasing absolute value of anisotropy parameter $\gamma$, quantum correlations decreases to a minimum value first and then increase to maximum value. Furthermore, the value of QD and TDD are nearly a constant when $\gamma$ is large enough as shown in Fig. 5. The constant is not the same for different temperature.

\section{Conclusion}
In summary, we have investigated the quantum correlations (quantum discord and trace distance discord) of the spin-$\frac{1}{2}$ Ising-XYZ chain on diamond structure. By changing the magnetic field, anisotropy parameter, coupling constant and also temperature we observed that the overall behavior of QD and TDD is alike to a large extent. We found that, the anisotropy parameter can both decrease and increase quantum correlation. Moreover, we observed regimes where QD and TDD increase by increasing temperature $T$, while concurrence decreases with $T$. Finally, we observed strong magnetic field suppresses quantum correlations, while weak magnetic field can increase or decrease QD and TDD. By comparison between QD and TDD versus magnetic field we showed that TDD is always greater than QD.

\section{Acknowledgement}
This research has been supported by Azarbaijan Shahid Madani university.

\end{document}